%% file: main.tex
\documentclass[10pt,twocolumn,letterpaper]{article}

\usepackage[accsupp]{axessibility}
\usepackage{cvpr}              
\usepackage{pifont}
\usepackage{marvosym}
\input{preamble}

\definecolor{cvprblue}{rgb}{0.21,0.49,0.74}
\usepackage[pagebackref,breaklinks,colorlinks,citecolor=green]{hyperref}

\makeatletter
\def\blfootnote{\xdef\@thefnmark{}\@footnotetext}
\makeatother

\def\paperID{1166} 
\def\confName{CVPR}
\def\confYear{2024}

\title{Bidirectional Autoregressive Diffusion Model for Dance Generation}

\author{Canyu Zhang\textsuperscript{1},
~Youbao Tang\textsuperscript{2}\textsuperscript{\Letter},
~Ning Zhang\textsuperscript{2},
~Ruei-Sung Lin\textsuperscript{2},\\
~Mei Han\textsuperscript{2}, 
~Jing Xiao\textsuperscript{3}, 
~Song Wang\textsuperscript{1}\textsuperscript{\Letter}\\
\textsuperscript{1}University of South Carolina, USA,
\textsuperscript{2}PAII Inc., USA,
\textsuperscript{3}Ping An Technology, China
}

\begin{document}
\maketitle

\blfootnote{\noindent This work was done during Canyu's intership in PAII Inc..
\textsuperscript{\Letter}Corresponding author.}

\begin{abstract}

Dance serves as a powerful medium for expressing human emotions, but the lifelike generation of dance is still a considerable challenge.
Recently, diffusion models have showcased remarkable generative abilities across various domains. 
They hold promise for human motion generation due to their adaptable many-to-many nature.
Nonetheless, current diffusion-based motion generation models often create entire motion sequences directly and unidirectionally, lacking focus on the motion with local and bidirectional enhancement.
When choreographing high-quality dance movements, people need to take into account not only the musical context but also the nearby music-aligned dance motions.
To authentically capture human behavior, we propose a Bidirectional Autoregressive Diffusion Model (BADM) for music-to-dance generation, where a bidirectional encoder is built to enforce that the generated dance is harmonious in both the forward and backward directions. To make the generated dance motion smoother, a local information decoder is built for local motion enhancement.
The proposed framework is able to generate new motions based on the input conditions and nearby motions, which foresees individual motion slices iteratively and consolidates all predictions.
To further refine the synchronicity between the generated dance and the beat, the beat information is incorporated as an input to generate better music-aligned dance movements. 
Experimental results demonstrate that the proposed model achieves state-of-the-art performance compared to existing unidirectional approaches on the prominent benchmark for music-to-dance generation.
Project page: \url{https://czzhang179.github.io/badm.github.io/}.

\end{abstract}

\begin{figure*}[t]
\centering
  \includegraphics[width=0.9\textwidth]{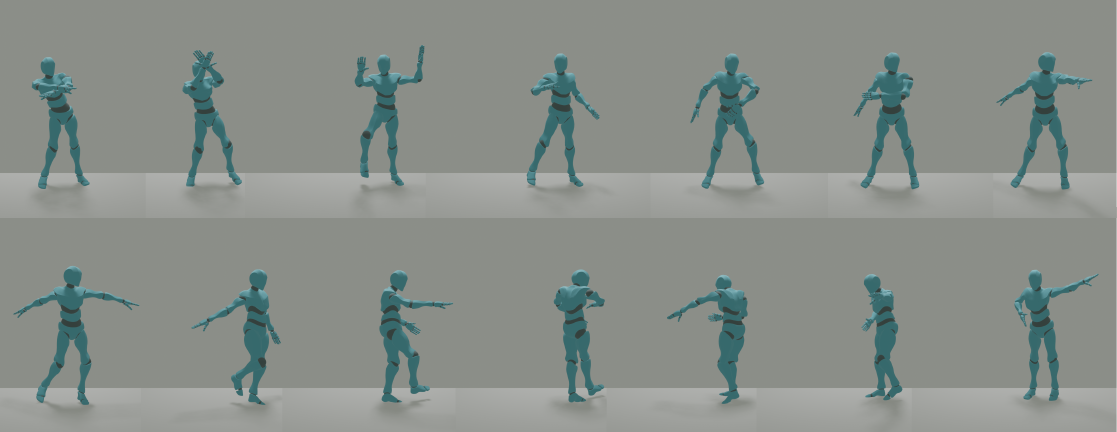} 
  \caption{Proposed bidirectional Autoregressive diffusion model (BADM) generates harmony, physically plausible dance based on music and beat conditions.}
  \vspace{-10pt}
  \label{fig:intro}
\end{figure*}

\section{Introduction}

Dance is a highly effective medium for expressing emotions, facilitating communication, and fostering social interaction. Nevertheless, the creation of new dances remains a formidable challenge, given the inherently expressive and freeform nature of dance movements.
In order to generate new dance sequences that seamlessly integrate with both preceding motions and the musical context, various deep learning-based methods have been proposed~\cite{kim2022brand, li2021ai, Gong_2023_ICCV, Li_2023_ICCV, Kim_2022_CVPR, marchellus2023m2c}. However, they typically treat dance generation as a matching problem. 
For example, some works~\cite{Zhang_2024_CVPR, siyao2022bailando, Yi_2023_CVPR} rely on constructing a codebook for music-to-motion matching, which hinders their potential for creative dance generation.
And they exclusively rely on past movements as the sole guide, neglecting the insights offered by future distributions.
These methods also encounter limitations when attempting to generate dances from music based on user-defined constraints.

Currently, diffusion models have shown remarkable promise in image processing tasks. 
Notable initiatives~\cite{tseng2022edge, tevet2022human} have explored the potential of diffusion models in the context of human motion generation. However, a significant drawback is the insufficient attention given to inter-frame transitions, resulting in sequences that lack coherence. 
When designing new motions, people always need to consider the music condition and nearby motions within a range.
But these models primarily emphasize conditioning input and global relationships, often neglecting the intricate details crucial for crafting fluid and harmonious motion.
Consequently, the generated dance movements for each frame may not align smoothly with the motions in nearby frames, leading to inconsistencies in the overall sequence.

In response to the limitations inherent in current methodologies, we present the innovative bidirectional Autoregressive diffusion model (BADM). 
Our model introduces a novel strategy by breaking down the entire noise sequence into smaller, manageable slices. 
During the generation of each slice, our model considers the preceding dance sequences and forwarding noise distributions using a cross-attention layer in a bidirectional way.
Then refined noise slices are sent into the decoder along with conditions iteratively.
The output dance slices are then concatenated and refined by a local information decoder to ensure a cohesive and harmonious overall sequence from a local perspective.

Another observation is that the beat information plays an important role in the art of dance generation, as it guides the timing of impressive movements.
Prior methodologies such as EDGE~\cite{tseng2022edge} exclusively rely on music features. 
To rectify this limitation, we extract the beat information as a distinct and independent condition.
Those features are fused with diffusion timestep as the whole condition.
Furthermore, we also segment both the music and beat features into corresponding slices, allowing the model to focus solely on each slice. 
This iterative approach ensures a more precise alignment with the rhythm and musical dynamics.

Our approach also offers exceptional editing capabilities ideally suited for dance choreography. 
It encompasses joint-wise conditioning and the ability to seamlessly interpolate between movements.
BADM also gains a remarkable ability to generate sequences of arbitrary length, granting it a high degree of versatility and adaptability.
In summary, our contributions are the following:

\begin{itemize}[itemsep=2pt,topsep=0pt,parsep=0pt,leftmargin=*]

    \item We propose a bidirectional autoregressive diffusion model based framework (BAMD) for music-to-dance generation. BADM first considers each motion slice separately and then refines them by utilizing their bidirectional dependencies and the local motion enhancement, so as to generate more harmony dances.  
    
    \item In order to enhance the beat information and generate better music-aligned dances, our model takes music beat as an independent condition. The music and beat information are segmented to help the model focus on each dance slice iteratively.

    \item Comprehensive experiments on the widely utilized music-to-dance datasets AIST++~\cite{li2021learn} demonstrate that our proposed method surpasses previous models by a significant margin across various metrics.

\end{itemize}

\begin{figure*}[t]
\centering
  \includegraphics[width=.85\textwidth]{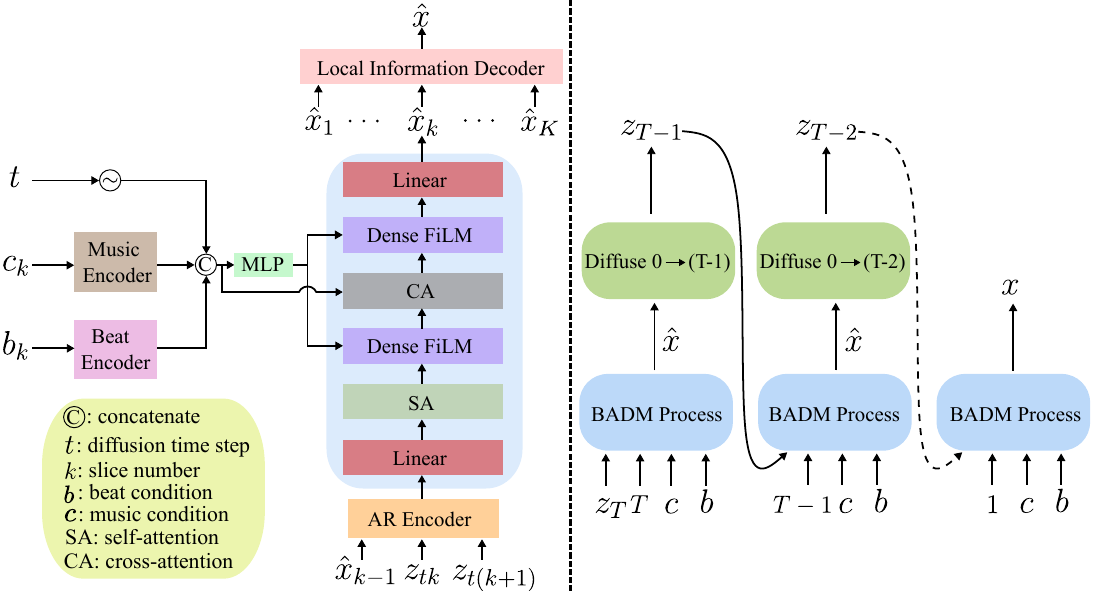} 
  \caption{Our bidirectional Autoregressive diffusion model (BADM) employs a denoising mechanism to enhance dance sequences from time $t = T$ to $t = 0$. BADM begins with a noisy sequence $z_T$ at time $T$, and proceeds to generate an estimated dance sequence $\hat{x}$. The denoising procedure is iteratively applied until $t=0$.
  Our autoregressive (AR) model treats the whole noise sequence as $K$ slices. On the left, we show our model processing the $k$-th slice at the diffusion timestep $t$. BADM is employed $K$ times within each BADM process.
  }
  \vspace{-10pt}
  \label{fig:main}
\end{figure*}

\section{Related Work}
\subsection{Human Motion Generation}

The quest to achieve lifelike human motion generation has been a longstanding pursuit.
Many prevalent approaches~\cite{arikan2002interactive, kovar2023motion, lee2002interactive} are rooted in graph-based methods, where motion sequences are decomposed into discrete nodes and then reassembled following predefined rules.
In recent years, deep neural networks have emerged as a promising alternative avenue for generating human motion. Some methods focus on predicting motion sequences based on initial pose sequences~\cite{hernandez2019human, guo2023back}, while others~\cite{kaufmann2020convolutional, harvey2020robust, duan2021single} employ bidirectional GRU and Transformer architectures for tasks like in-betweening and super-resolution.
For instance, Holden et al.\cite{holden2016deep} utilize autoencoder to acquire a latent representation of motion, which is then used to manipulate and control motion with respect to spatial factors such as root trajectory and bone lengths.
Moreover, motion generation can be guided at a higher level by external cues, including action classes~\cite{cervantes2022implicit}, audio signals~\cite{aristidou2022rhythm}, and natural language descriptions~\cite{petrovich2022temos}.
Notably, Tevet et al.~\cite{tevet2022human} harnesses the power of pre-trained large language models like CLIP \cite{radford2021learning} to establish a shared latent space for both language and motion.

\subsection{Music-to-dance Generation}

The challenge of generating dances that synchronize with the input music has fueled the development of a wide array of deep learning-based techniques. 
These methods span a diverse spectrum, encompassing CNN~\cite{holden2016deep}, GANs~\cite{sun2020deepdance, lee2019dancing}, and Transformer models~\cite{li2020learning, li2021ai}. Their common goal is to directly translate the provided music into a continuous sequence of human poses.
For instance, Esser et al.~\cite{esser2021taming} encodes intricate visual elements into quantized patches, employing Transformers to generate contextually coherent images at high resolutions, bridging the visual and musical aspects of dance generation.
In a different approach, Bailando~\cite{siyao2022bailando} adopts VQ-VAE~\cite{van2017neural} to maintain temporal coherence across various music genres, a crucial element in dance generation, ensuring that the movements align with the music's rhythm and style.
And EDGE~\cite{tseng2022edge} leverages conditional diffusion models to craft human dance movements in direct response to musical cues, utilizing the potent audio feature extractor Jukebox~\cite{dhariwal2020jukebox}. Diffusion based models also support human motion editing similar as inpainting~\cite{zhang2023sair, zhang2023superinpaint, li2022misf, guo2021jpgnet}.

\subsection{Diffusion Model}

Diffusion models~\cite{ho2020denoising, sohl2015deep} belong to the realm of neural generative models. They draw inspiration from the stochastic diffusion process, a concept in $thermodynamics$. In this paradigm, data distribution samples undergo a gradual introduction of noise through the diffusion process. 
Subsequently, a neural model is trained to reverse this process, gradually restoring the original sample to its pristine state. Sampling from the learned data distribution involves denoising an initially pure noise vector.
The evolution of diffusion models in the field of image generation has been enriched by earlier works~\cite{ho2020denoising, song2020denoising}. For conditioned generation tasks, Dhariwal et al.~\cite{dhariwal2021diffusion} introduce classifier-guided diffusion, a concept later embraced and extended by GLIDE~\cite{nichol2021glide}, which enables conditioning based on CLIP textual representations.
In recent developments, some works~\cite{petrovich2022temos, kim2023flame, tevet2022human, lee2022autoregressive, karunratanakul2023guided, li2023finedance} have advocated for the adoption of diffusion models in the context of motion generation tasks and earn promising results.

\subsection{Autoregressive Model}

An autoregressive model is a statistical or machine learning paradigm designed to analyze and predict sequential data. It operates by making predictions for each data point based on its dependencies on previous observations within the sequence. In essence, it posits that each data point's value is influenced by the values of its predecessors.
While autoregressive models have found wide-ranging applications in fields such as image processing~\cite{wei2023autoregressive, huang2023towards, maiya2023nirvana}, they have been relatively underutilized in the domain of motion generation~\cite{wen2021autoregressive}. 
Given the unique characteristics of dance generation, we introduce an autoregressive diffusion model to address this specific task.

\section{Method}


\subsection{Pose Representation}

Our objective is to generate a human motion sequence with a length of $N$, given arbitrary musical condition $c$ and beat condition $b$.
We represent these dance sequences as a series of poses using the 24-joint SMPL format~\cite{loper2023smpl}. Each joint's orientation is represented using a 6-DOF rotation representation~\cite{zhou2019continuity}, and a single translation vector for the root, resulting in a total pose vector $p \in \mathbb{R}^{24 \times 6+3=147}$.
Following previous work EDGE~\cite{tseng2022edge}, we incorporate binary contact labels for the heel and toe of each foot, resulting in a binary contact label vector $l \in \mathbb{R}^{2 \times 2=4}$.
Consequently, the complete representation of the pose sequence is  $ x \in \mathbb{R}^{N \times 151}$.

\subsection{Diffusion Framework}

Our diffusion process is represented as a Markov noise process~\cite{ho2020denoising}.
We generate the noise sequence $\{z_t\}^T_{t=0}$ for each timestep $t$.  In the forward noising process, $x \sim p(x)$ is initially sampled from the data distribution. 
The forward noise process is defined as follows:
\begin{align}\label{eq:importance2}
q(z_t | x) = \mathcal{N}(\sqrt{\alpha_t}x, (1-\alpha_t)I),
\end{align}
$\alpha_t \in (0,1)$ represents constant hyper-parameters. When $\alpha_t$ is sufficiently small, we can approximate the output as $z_T \sim \mathcal{N}(0, 1)$. 

To recover the clean dance sequence, our conditioned diffusion motion model treats the distribution $\hat{x}_{\theta}(z_t, t, c, b) \approx x$ as the reverse diffusion process for gradually cleaning $z_t$, with model $\theta$ for all diffusion timestep $t$. 
Instead of predicting the variation $\epsilon_t$ as formulated by DDPM ~\cite{ho2020denoising},  we predict the signal itself, using the following simple objective:
\begin{align}
\mathcal{L}_{simple} = \mathbb{E}_{x, t}[||x -\hat{x}_{\theta}(z_t, t, c, b)||_2^2].
\end{align}

\subsection{Geometric Losses}

In addition to the reconstruction loss $\mathcal{L}_{simple}$, we adopt geometric auxiliary losses similar to HDM~\cite{tevet2022human} and EDGE~\cite{tseng2022edge}, 
which encourage the matching in three  aspects of physical realism: 
joint positions ($\mathcal{L}_{pos}$, \ref{pos}), velocities ($\mathcal{L}_{vel}$, \ref{vel}), and foot contact ($\mathcal{L}_{foot}$, \ref{contact}).
These losses enforce physical properties
and prevent artifacts, encouraging natural and coherent motions

\begin{align}
\mathcal{L}_{pos} = \frac{1}{N} \sum\limits_{i=1}^N || FK(x^i) - FK(\hat{x}^i)||_2^2 ,
\label{pos}
\end{align}

 \begin{align}
\mathcal{L}_{vel}  = \frac{1}{N-1} \sum\limits_{i=1}^{N-1} || (x^{i+1}-x^{i}) - (\hat{x}^{i+1} - \hat{x}^{i})||_2^2,
\label{vel}
\end{align} 

\begin{align}
\mathcal{L}_{foot} = \frac{1}{N-1} \sum\limits_{i=1}^{N-1}  || (FK(\hat{x}^{i+1}) - FK(\hat{x}^i) \cdot f_i)||_2^2,
\label{contact}
\end{align}
the function $FK(\cdot)$ represents the forward kinematic process, which translates joint rotations into joint positions, and the variable $i$ is the frame index.
In $\mathcal{L}_{foot}$,  $f_i$ signifies the model's internal estimation of the binary foot contact label's influence on the pose at each frame. 
This approach not only incentivizes the model to forecast foot contact accurately but also enforces it to maintain coherence with its self-generated predictions.
Overall, our training loss is the weighted sum of the simple objective and the geometric losses: 
\begin{align}
\mathcal{L} = \mathcal{L}_{simple}  + \lambda_{pos}\mathcal{L}_{pos}  + \lambda_{vel}\mathcal{L}_{vel} + \lambda_{foot}\mathcal{L}_{foot}.
\end{align}

\subsection{Model}

\begin{figure}[t]

\centering
  \includegraphics[width=.4\textwidth]{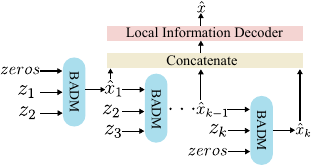} 
  \caption{BADM processes each noise slice $z_k$ in a bidirectional way. Generated dance slices are concatenated and sent to the local information decoder. We show this process at each timestep t.}
  \vspace{-10pt}
  \label{fig:main1}
\end{figure}

Our model architecture is depicted in Figure~\ref{fig:main} and Figure~\ref{fig:main1}. 
The inputs of our model are noise slice $z_t$, diffusion timestep $t$, music condition $c$, and beat condition $b$, where $z_t$, $c$, and $b$ have the same length.
To ensure that the newly generated motion remains faithful to the preceding and future motions, we segment the entire noise sequence into smaller slices. 
Our model takes into consideration the previous motions and future noise distributions when generating the new motion slice.
To elaborate on the process, we divide the complete noise sequence $z_t$ into $K$ slices at the timestep t, represented as $(z_{t1}, z_{t2}, ..., z_{tK})$. Each noise slice $z_{tk}$ undergoes a cross-attention layer that considers the previously generated dance sequence $
\hat{x}_{(k-1)}$ and the subsequent noise part $z_{t(k+1)}$.
Each processed noise slice $\hat{z}_{tk}$ is subsequently sent into the decoder, which contains feature-wise linear modulation (FiLM)~\cite{perez2018film}, alongside the segmented music feature $c_k$ and beat information $b_k$,
the output is the generated dance slice $\hat{x}_{k}$.

We also include zero padding for the initial and final parts to ensure a smooth transition and maintain the integrity of the generated dance motions.
To ensure temporal context is preserved, we incorporate timestep information through a token that is concatenated with the music conditioning.
The resulting dance sequences $(\hat{x}_{1}, \hat{x}_{2}, \dots, \hat{x}_{K})$ are concatenated and sent into a local information decoder. 
The local information encoder is constructed using 1D convolutional layers.
This step is essential to ensure that the motions from nearby frames exhibit harmony and cohesion. 
The final output $\hat{x}$ is the generated dance which is faithful to music and beat conditions.

\subsection{Sampling}

At each denoising timestep $t$, BADM predicts the denoised sample, then reverberates the effect back to timestep $t - 1$ as follows: $\hat{z}_{t-1} \sim q(\hat{x}_\theta(\hat{z}_{t}, c, b), t-1)$, ultimately concluding this process when $t$ reaches zero.
Following previous works, our model's training approach utilizes classifier-free guidance~\cite{ho2022classifier}. 
This strategy is implemented by randomly substituting the conditioning variable with $\{c, b\} = \emptyset$ during training, albeit with low probability.
The outcome of guided inference is a blend of unconditionally generated samples and conditionally generated samples, expressed as a weighted summation:
\begin{align}
\hat{x}(\hat{z}_t, c, b) = w \cdot\hat{x}(\hat{z}_t, c, b)
+ (1-w)\cdot\hat{x}(\hat{z}_t, \emptyset).
\end{align} 
We can amplify the conditions by setting guidance weight $w \textgreater 1$ during sampling.

\subsection{Long-form Sampling}

The capability to create dance sequences of diverse lengths, even spanning several minutes, stands as a vital necessity in dance generation. 
The simple idea is to increase the length of the condition input,
but it can result in a linear escalation of computational demands as sequences grow in length. 
Additionally, the conditioning factors $c$ and $b$ can exert additional pressure on memory resources.

In order to produce extended sequences, we employ a methodology based on segmenting the sequences into $N$-frame slices. 
To maintain uniformity across adjacent $N/2$-frame slices, we apply interpolation, employing a linearly decreasing weight. 
Since the frames within each slice have been encoded through an autoregressive encoder, we contend that the entire slice inherently encapsulates valuable autoregressive information.

\subsection{Editing}

To facilitate the post-processing of dances generated by BADM, we employ a well-established masked denoising technique inspired by previous models~\cite{lugmayr2022repaint, tevet2022human}.
Our model offers the flexibility to accommodate a variety of constraints, allowing users to apply both temporal and joint-wise specifications. 
When provided with a joint-wise or temporal constraint $x^{known}$ with positions indicated by a binary mask $m$, the following operations are performed at each denoising timestep:
\begin{align}
\hat{z}_{t-1} := m \odot q(x^{known}, t-1)
+ (1-m)\odot\hat{z}_{t-1},
\end{align} 
where $\odot$ refers to the Hadamard product, which operates element-wise to replace the regions of prior knowledge with forward-diffused samples of the constraint. This technique ensures that the dance sequences remain editable during the inference phase, without requiring any additional training processes.

\subsection{Implement Details}

We draw inspiration from recent advancements and utilize Jukebox features as the music condition input. 
The research in music information retrieval~\cite{castellon2021codified} has shown that pre-trained Jukebox~\cite{dhariwal2020jukebox} model can generate highly effective representations for audio tasks. 
However, the jukebox feature doesn't contain enough beat information, which can be reflected in generated motions directly.
To augment the rhythmic aspect, we incorporate one-hot encoding of the music beat as a condition, which is extracted by the public audio processing toolbox Librosa~\cite{jin2017towards}.

In our experiments, we set the hyperparameter $K=6$.
When training our models, we employ the Adan optimizer with a learning rate of 0.0002. All models undergo training for 2000 epochs, utilizing four NVIDIA Tesla V100 GPUs for a period of about 1 day, and the batch size is set to 128.




\begin{table*}[t]
\centering
\renewcommand\arraystretch{1.2}
\footnotesize
 \resizebox{0.85\textwidth}{!}{
\begin{tabular}{l|ccccccccc}
\toprule
Method  &  $\text{Dist}_k\rightarrow$  & $\text{Dist}_g\rightarrow$  & BA $\uparrow$ &  PFC$\downarrow$ &Ours Win Rate& Fixed bones & Editing\\  
\midrule
 Ground Truth&8.19& 7.45&0.2374&1.332 & $66.6\% \pm 10.0\%$ & \ding{51} &  N/A \\
\midrule
FACT~\cite{li2021learn} (ICCV2021)&5.94& 6.18& 0.2209& 2.254 &$91.6\% \pm 5.0\%$ & \ding{51} &  \ding{55} \\
Bailando~\cite{siyao2022bailando} (CVPR2022)& 7.83 & 6.34 &0.2332&1.754 &$86.6\% \pm 6.6\%$& \ding{55} & \ding{55}\\
EDGE~\cite{tseng2022edge} (CVPR2023)& 9.48 &5.72& 0.2281&1.654&$64.9\% \pm 8.3\%$ & \ding{51} & \ding{51}  \\
\midrule
BADM (ours) &\textbf{8.29}&\textbf{6.76}&\textbf{0.2366}&\textbf{1.424} & N/A & \ding{51} & \ding{51}\\
\bottomrule
\end{tabular}}

 \caption{We compare BADM against FACT~\cite{li2021learn}, Bailando~\cite{siyao2022bailando}, and EDGE~\cite{tseng2022edge}. 
$\uparrow$ means higher is better, $\downarrow$ means lower is better, and $\rightarrow$ means closer to ground truth is better. We obtain the quantitative results from their respective publications~\cite{siyao2022bailando, tseng2022edge}, or re-evaluation results using published code.
}
\vspace{-10pt}
\label{main-result}
\end{table*}

\section{Experiments}

\subsection{Dataset}
In our study, we leverage the AIST++ dataset~\cite{li2021learn}, which is composed of  1,408 meticulously curated dance motions intricately synchronized with music spanning a wide spectrum of genres. 
We adhere to the train/test partitioning scheme established by the dataset creators. 
To maintain consistency and facilitate analysis, all training examples have been truncated to a uniform duration of 5 seconds, captured at 30 frames per second (FPS).

\subsection{Baselines}

We have selected FACT~\cite{li2021learn}, Bailando~\cite{siyao2022bailando}, and EDGE~\cite{tseng2022edge} as the baselines for our study.
FACT~\cite{li2021learn} is a full attention cross-modal transform model that can generate a long sequence of
realistic 3D dance motion.
Bailando~\cite{siyao2022bailando} is a subsequent approach, demonstrating remarkable qualitative performance improvements.
Lastly, EDGE~\cite{tseng2022edge} is the most recent transformer-based diffusion model dance generation. They are the state-of-the-art traditional and diffusion based music-to-dance generation models.

\subsection{Results}

\paragraph{Generation diversity.}
To evaluate our model’s ability to generate diverse dance motions when given various
input music, we compute the average feature distance (DIV) in the feature space as proposed in~\cite{siyao2022bailando, li2020learning}.  
We evaluate the generated dances in two feature spaces: kinetic feature~\cite{onuma2008fmdistance} (denoted as 'k') and geometric feature~\cite{muller2005efficient} (denoted as 'g').
The quantitative results are presented in Table~\ref{main-result}. 
Our method outperforms EDGE in terms of both $\text{Dist}_k$ and $\text{Dist}_g$, achieving improvements of 1.19 and 1.04 in each metric. 
Furthermore, when compared with Bailando, BADM demonstrates a remarkable enhancement of 0.46 in $\text{Dist}_k$,
showcasing its ability to generate diverse choreographies instead of converging to a limited set of templates. 
The BADM autoregressive encoder takes into account only the motions in nearby frames, providing it with the flexibility to adapt to the specific context of each frame.

\paragraph{Motion-music correlation.}
To determine how well the generated dance sequences align with the accompanying music, we calculate the Beat Align Score (BA) following ~\cite{siyao2022bailando}. This score quantifies the average temporal distance between each beat in the music and its nearest corresponding beat in the dance sequence.
As shown in  Table~\ref{main-result}, BADM beats all previous methods on this metric.
These findings also highlight BADM's proficiency in improving the correlation between music and motion.

\paragraph{Physical plausibility.}
The Physical Foot Contact score (PFC) ~\cite{tseng2022edge}  evaluates the plausibility of foot-ground interaction, without presuming that the feet should maintain static contact throughout the entire dance sequence.
As shown in  Table~\ref{main-result}, BADM demonstrates a substantial improvement of 0.23 compared to EDGE.
And BADM boosts the PFC performance by 0.67 compared to Bailando. 
These results signify BADM's superior physical plausibility, driven by the local information decoder, which allows the model to focus on the motion at each frame.

\paragraph{Motion quality.}
Some dance generation works~\cite{siyao2022bailando, li2021learn} use the Fréchet Inception Distance (FID) metric~\cite{hernandez2019human} to evaluate the quality of generated dance sequences. FID is employed to quantify the dissimilarity between the generated dance sequences and the entirety of motion sequences present in the AIST++ dataset, encompassing both the training and test data.
Nevertheless, the reliability of this metric has been called into question by EDGE~\cite{tseng2022edge}, based on their insightful observations. 
EDGE has pointed out potential issues, particularly concerning the limited coverage of the AIST++ test set given its relatively small size.
In spite of these concerns, our model (20.24) still outperforms EDGE (24.71) when assessed using the $\text{FID}_g\downarrow$ metric, which means BADM can generate the dance with high quality.

\paragraph{User study.}
To gain a deeper understanding of the true visual performance of our method, we conducted a comprehensive user study, comparing the dance sequences generated by our approach with the baselines on AIST++ dataset. This study involved 12 participants, each participating in individual evaluations.
Each participant was presented with a series of 30 pairs of comparison videos, each lasting approximately 10 seconds. 
These pairs included our generated results and those generated by one of our competing methods, all synchronized to the same music track. 
Participants were tasked with discerning which of the two videos exhibited better synchronization with the music.

The detailed statistics derived from this user study can be found in Table~\ref{main-result}. 
Notably, our method significantly surpasses the compared state-of-art method EDGE with 64.9\% winning rates. 
Our method even has a 66.6\% rate to beat the ground truth dance.
This achievement is particularly remarkable given that the baseline dances exhibited noticeable distortions during intricate movements.

\begin{table}[t]
\centering
\renewcommand\arraystretch{1.2}
\footnotesize
 \resizebox{0.35\textwidth}{!}{
\begin{tabular}{l|cccc}
\toprule
Method   &  Ours Win Rate  \\  
\midrule
Bailando~\cite{siyao2022bailando} (CVPR 2022) & $83.3\% \pm 6.7\%$ \\
EDGE~\cite{tseng2022edge} (CVPR 2023) & $61.6\% \pm 8.3\%$\\ 
\midrule
BADM (ours) & N/A \\
\bottomrule
\end{tabular}}

 \caption{User study results from in-the-wild music. We compare BADM against Bailando~\cite{siyao2022bailando} and EDGE~\cite{tseng2022edge}.}
 \vspace{-10pt}
\label{in-the-wild}
\end{table}

\begin{figure*}[th]
\centering
  \includegraphics[width=.9\textwidth]{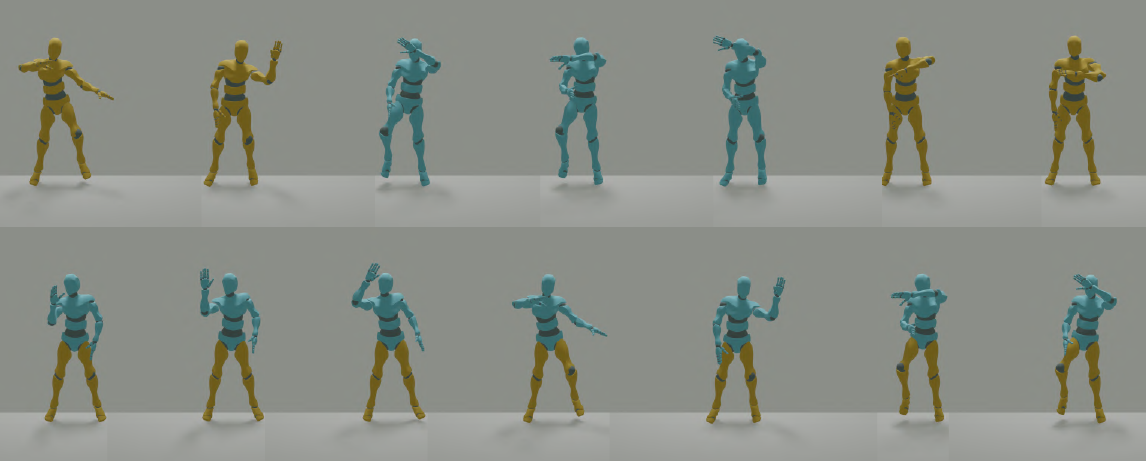} 
  \caption{Yellow parts represent fixed motion inputs and blue parts are the generated motion parts. For motion in-betweening (Top), the first and last frames are fixed. For specific body part editing (Bottom), the lower body joints are fixed to the input motion while the upper body is altered to fit the input.}
  \vspace{-10pt}
  \label{fig:editing}
\end{figure*}

\paragraph{In-the-wild music.}
While our method has showcased impressive performance on the AIST++ dataset, it is crucial to acknowledge that these achievements may not automatically extend to the 'in-the-wild' music inputs.
To evaluate this critical aspect of generalization, we conducted an assessment of the proposed method and the baseline approaches. 
We select several popular songs from YouTube for this evaluation.
As shown in Table~\ref{in-the-wild}, BADM has an 83.3\% rate to win the Bailando, and a 61.6\% rate to beat EDGE.
Those results reflect that our model succeeds in generating the dances based on diverse music inputs.

\paragraph{Qualitative results.}

We present several qualitative examples in Figure~\ref{fig:intro}, where our model's capabilities are on full display. Our model excels in generating harmonious dance sequences that seamlessly incorporate both footwork and hand movements.
In contrast, previous methods, such as EDGE~\cite{tseng2022edge}, often falter during complex movements. 
For instance, they may produce inaccuracies, causing the generated human body to be 'lying down' when it should be 'turning around'.

\subsection{Additional Study}

\paragraph{Fixed bones.}
Preserving fixed bone length is a fundamental criterion for assessing physical plausibility. Our approach excels in this aspect by operating within the reduced coordinate space (joint angle space), enabling us to consistently maintain bone lengths. In contrast, methods that operate in the joint Cartesian space, often result in substantial variations in bone lengths. For example, Bailando can fluctuate bone lengths by as much as 20\%.

\paragraph{Motion editing.}
In this section, we introduce two motion editing applications: 'in-between' and 'body part editing,' each following certain restrictions in the temporal and spatial domains, respectively.
In the 'in-between' application, we maintain the initial and final motions as fixed reference parts and challenge the model to generate the intermediate motions.
In the case of 'body part editing,' we identify the joints to remain unaltered, allowing the model to autonomously create the desired adjustments. 
In our experiments, we focus on exploring the possibilities of exclusively editing the upper body joints.
As showcased in Figure~\ref{fig:editing} (TOP), our method demonstrates its proficiency in generating fluid and coherent motion sequences that seamlessly bridge the initial and final motions.
Furthermore, as depicted in Figure~\ref{fig:editing} (Bottom), our method excels in generating upper body motions in sync with the music while keeping the lower body fixed.

\begin{table}[t]
\centering
\renewcommand\arraystretch{1.2}
\footnotesize
 \resizebox{0.47\textwidth}{!}{
\begin{tabular}{l|cccc}
\toprule
Method   &  $\text{Dist}_k\rightarrow$  & $\text{Dist}_v\rightarrow$  & BA $\uparrow$ &  PFC$\downarrow$ \\  
\midrule
 GT&8.19& 7.45&0.2374&1.332  \\
\midrule

$K=3$ &8.49&6.34&0.2290& 1.655 \\
$K=5$ &8.38&6.68&0.2338& 1.482\\
$K=10$  &7.75&5.82&0.2189&1.756 \\
\midrule
w/o Beat&7.93&6.62&0.2235&1.537 \\
Unidirection&7.86&6.46&0.2339&1.582\\

Removing LID&8.02 &6.60&0.2319&1.542\\
\midrule
BADM &\textbf{8.29}&\textbf{6.76}&\textbf{0.2366}&\textbf{1.424}\\
\bottomrule
\end{tabular}}

 \caption{Ablation study results from beat information, segmentation step, and unidirectional encoder. $\uparrow$ means higher is better, $\downarrow$ means lower is better, and $\rightarrow$
means closer to ground truth is better.}
\label{ablation-result}
\vspace{-10pt}
\end{table}

\subsection{Ablation Study}

\paragraph{Beat information.}

In our setup, we introduce one-hot beat information as a conditioning factor, and we also explored the model's performance when this condition is removed.
As illustrated in Table~\ref{ablation-result}, the addition of the beat condition yields a notable improvement in the beat alignment score, showcasing an increase of 0.0131. 
This outcome underscores the importance of this condition, as it provides the model with more precise and specific guidance for generating dance sequences.

\paragraph{Segmentation step size.}

In our experiments, we conducted a comprehensive analysis of different choices for the hyperparameter $K$. 
Initially, we set $K=6$, and we explored various alternatives. 
As summarized in Table~\ref{ablation-result}, the results unequivocally favor the model with $K=6$ as the top-performing configuration.
Specifically, the model with $K=6$ surpasses its counterpart with $K=3$ by a significant margin of 0.2 on the $\text{Dist}_k$ metric. 
Additionally, our model outperforms the model with $K=10$ by 0.332 on the PFC metric. 
An optimal choice of $K$ can lead to a substantial increase in performance.
When $K$ is excessively large, each time slice encompasses multiple movements, bringing too many restrictions to the model. 
Conversely, when $K$ is too small, individual movements are segmented into excessively meaningless parts, limiting the information available for analysis.

\paragraph{Unidirectional encoder.}

In our experimental setup, the autoregressive encoder incorporates a bidirectional approach, allowing it to consider the input from both forward and backward directions. We also conducted experiments with a unidirectional encoder.
In the unidirectional encoder configuration, each noise slice $z_{tk}$ undergoes a cross-attention layer that only takes into account the dance sequence $\hat{x}_{(k-1)}$ generated up to that point. 
The results are depicted in Table~\ref{ablation-result}.
The bidirectional approach can increase the PFC by 0.158 and boost the $\text{Dist}_k$ metric by 0.43.
Those results reflect that incorporating both forward and backward information is pivotal in enabling the model to generate dance sequences that are harmonious and consistent. 
It's worth mentioning that the noise sequence gradually aligns with the original dance, as the added noises become progressively smaller.

\paragraph{Local information decoder.}

In our model, we leverage 1D convolution layers to enhance the intricacies of the dance sequence from a local perspective. The impact of this refinement is evident in the outcomes presented in Table~\ref{ablation-result}. Notably, when the local information decoder is omitted, both the PFC and $\text{Dist}_k$ experience a discernible decline, amounting to 0.118 in the former and a degradation of 0.27 in the latter.
These findings serve as compelling evidence supporting the efficacy of our chosen components.

\section{Discussion}

In the final stage of our proposed model, we employ a local information decoder to enhance the harmony of the generated motions. We also experimented with a global information encoder to refine the output. 
%
However, the results were less than satisfying. The spatial freedom of the body's hand and foot movements became constrained, and the body struggled to execute certain complex movements. 
Upon careful examination, we attribute this issue to the global information component, which compels the motion at each frame to account for actions in distant frames, ultimately restricting creative expression.
These results further validate our decision to generate each motion based on nearby motions rather than considering the entire sequence of motions.


\section{Future Work}

In our current model, we segment the entire dance sequence using a fixed step size. 
In reality, a dance performance consists of distinct lengths and meaningful movements, such as 'jumping', 'turning around', and 'pirouette'. 
Our existing segmentation method may inadvertently split these complete movements into separate segments, destroying the original meaning and disrupting the dance's fluidity. 
In contrast, when the segment length is too long, it might encompass multiple independent movements.
In our future research, we aim to enhance our segmentation technique by taking into account the semantic meaning of each movement. 
This means that we will strive to identify and segment the dance sequence in a way that respects the integrity of individual movements, ensuring a more accurate representation of the dancer's artistry and choreography.

\section{Conclusion}

In our research, we introduce a bidirectional autoregressive diffusion model (BADM) to tackle the challenge of music-to-dance generation. 
Recognizing the shortcomings of prior approaches, we have devised an autoregressive encoder specifically tailored for processing dance slices. 
This novel encoder allows our model to create motion at each frame while taking into account the neighboring frame motions in a range, resulting in generating smoother and more coherent dance sequences.
And we propose the local information decoder to further refine the final prediction.
Moreover, we have enriched our model by incorporating beat information, which significantly enhances the connection between the generated movements and the underlying music. 
Through a series of comprehensive experiments and meticulous ablation studies, 
we have demonstrated the superior performance of our methods over existing techniques on the AIST++ dataset across various evaluation metrics.
Our research offers a new means to produce harmonious and lifelike dances that resonate with music.

{
    \small
    \bibliographystyle{ieeenat_fullname}
    \bibliography{main}
}


\end{document}


\maketitle
\section{Study on Pixel-wise Importance Map}

\begin{table}[h]\LARGE
	\centering
	
	\renewcommand\arraystretch{1.0}
	\footnotesize
	\setlength{\tabcolsep}{0.8mm}{
		\begin{tabular}{l|cccc}
			\toprule
			  
              Method &  PSNR$\uparrow$  & SSIM$\uparrow$  &   L1 $\downarrow$ &  LPIPS$\downarrow$   \\
			\midrule
		   Using binary mask  &28.26&0.942 &0.025 & 0.120\\
               Using PIM   & \textbf{29.84} & \textbf{0.968} & 
               \textbf{0.022} & \textbf{0.053}  \\
    
			\bottomrule
	\end{tabular}}
\vspace{10pt}
 \caption{Ablation study on pixel-wise importance map.}
\label{ablation-map}

\end{table}

\section{More Qualitative Results}


%% file: preamble.tex
\usepackage[dvipsnames]{xcolor}
